\documentclass[prl,aps,showpacs,twocolumn]{revtex4}
\usepackage{graphics,bm}
\usepackage{amssymb}
\usepackage{epsfig}
\usepackage{epsf}
\usepackage[usenames]{color}

\begin{document}

\title{Spin drag Hall effect in a rotating Bose mixture}

\author{H.J. van Driel}

\author{R.A. Duine}

\author{H.T.C. Stoof}

\affiliation{Institute for Theoretical Physics, Utrecht
University, Leuvenlaan 4, 3584 CE Utrecht, The Netherlands}
\date{\today}

\begin{abstract}
We show that in a rotating two-component Bose mixture, the spin
drag between the two different spin species shows a Hall effect.
This spin drag Hall effect can be observed experimentally by
studying the out-of-phase dipole mode of the mixture. We determine the damping of this mode due to spin drag as a
function of temperature. We find that due to Bose stimulation
there is a strong enhancement of the damping for temperatures
close to the critical temperature for Bose-Einstein condensation.
\end{abstract}


\maketitle

\def\bx{{\bf x}}
\def\bk{{\bf k}}

{\it Introduction}
---
Electronic transport is one of the main topics of interest in
condensed-matter physics, and an especially important phenomenon
in electronic transport is the Hall effect. It was discovered
already in the late nineteenth century by Hall. He observed that
if a magnetic field ${\bf B}$ is applied perpendicular to the
current density ${\bf j}$ through a conductor, the Lorentz force
leads to a voltage drop in the direction perpendicular to both the
current density and the magnetic field. This voltage is
proportional to ${\bf j} \times {\bf B}$, with a proportionality
constant that depends only on the density of electrons and not on
any other material parameters \cite{hall1879}. While the discovery
of the Hall effect predates that of the electron, it is important
for our purposes to note that electronic transport is in fact
fermionic transport since it is mediated by the movement of
electrons. By now many variations of the Hall effect have been
found: the (integer and fractional) quantum Hall effects in which
the Hall voltage is quantized
\cite{PhysRevLett.45.494,PhysRevLett.50.1395,PhysRevLett.48.1559},
the spin Hall effect
\cite{ShuichiMurakami09052003,PhysRevLett.92.126603}, and the quantum
spin Hall effect \cite{MarkusKonig09202007}. Recently, a spin Hall
drag effect has also been proposed \cite{PhysRevLett.103.196601}.
The latter is, as all spin Hall effects are, due to spin-orbit
interactions that play no role in atomic Bose gases if they are
not externally introduced by applying an appropriate laser field
configuration \cite{PhysRevA.78.023616}. The proposal of
Ref.~\cite{PhysRevLett.103.196601} is therefore physically very
different from what we discuss below.

An important field of physics which connects few-body atomic physics with many-body and condensed-matter physics is that of cold atoms. The
Bose-Einstein condensation of bosons at very low temperatures was
already predicted by Einstein in 1924, but only observed
directly in 1995 \cite{anderson1995}. Over the past fifteen years, techniques
have been getting steadily more refined, and it is now possible to
make all sorts of degenerate atomic mixtures consisting of several
spin states or of several different atomic species of either
fermions or bosons. These mixtures are always trapped in optical
and/or magnetic potentials, in which the atoms can be set into
rotation by stirring with a so-called laser spoon \cite{PhysRevLett.84.806}.

It is tempting to combine the above two fields to also get more
insight into the physics of bosonic transport. At first sight this
seems less than straightforward, since it is not possible to
simply attach leads to a cloud of cold atoms to set up a
steady-state transport current of atoms through the mixture. Moreover, in the
cold-atom situation there are no obvious mechanisms that relax the
particle current and give nonzero resistivities. However, we can
use the phenomenon of spin drag as a bridge between these two
worlds
\cite{polini2007,PhysRevLett.103.170401,PhysRevLett.104.220403}. Spin drag was first proposed by
D'Amico and Vignale by making an analogy with Coulomb drag between
to electron layers \cite{damico2000}. It was later observed by
Weber {\it et al.} \cite{weber2005}. Whereas in the classic
Coulomb drag experiment electrons are differentiated by the layer
they occupy, in spin drag the spin of the electron is the relevant
degree of freedom, i.e., electrons of one spin species drag along
electrons of the other spin species. The resistivity created by
this spin drag, which is a resistivity to spin but not to charge
currents, typically goes as $\rho_D \propto T^2$ in electronic
systems, with $T$ the temperature. This is the trademark of a
Fermi-liquid like behavior.

In our earlier work \cite{PhysRevLett.103.170401}, we investigated the situation in
which the particles involved in the spin drag are bosons instead
of fermions. We proposed an idealized set-up in which spin-1
bosons in the state $|m_F = +1\rangle$ are accelerated along a
torroidal trap by a time-dependent magnetic-field texture that creates a
fictitious electric field. The atoms in state $|m_F = 0\rangle$,
which are also present, do not feel this force but experience spin
drag due to collisions with the other species. We found that due
to the Bose enhancement of interatomic scattering, the drag
resisitivity increases at lower temperatures. For the
one-dimensional set-up considered, it in fact behaves as $\rho_D
\propto T^{-5/2}$ for low temperatures, in strong contrast with the
usual quadratic Fermi-liquid result.

In this Letter, we discuss spin drag in a rotating Bose mixture.
We consider the realistic situation of a three-dimensional Bose
mixture with two spin components, present in equal numbers, just
above the temperature for Bose-Einstein condensation. We first
consider the homogeneous case and look in linear response for
steady-state solutions of the appropriate Boltzmann equation with
a nonzero spin current. We find that the drag resistivity now
becomes a $3 \times 3$ matrix with nonzero off-diagonal elements
that are proportional to the rotation speed, which represents a
Hall effect. Indeed, these off-diagonal elements are analogous to
those found in the classical Hall effect and, in particular, do not
depend on the specific collisional details of the mixture that
determine the diagonal resistivities, but only on the atomic
density and external rotation frequency.

As mentioned previously, such steady-state solutions no longer
exist in the realistic situation that the atomic mixture is
trapped in an external harmonic potential. The spin drag Hall
effect can nevertheless be observed in that case by considering
the collective modes. In particular, we consider the dipole mode
in which the two spin components oscillate out of phase with each
other, because this mode obtains an orthogonal, i.e., a transverse
component due to the spin drag Hall effect. Moreover, the
longitudional spin drag leads to damping of this mode, which makes
it interesting to find out how the relaxation rate of these modes
depends on temperature. To obtain this, we again solve the
Boltzmann equation for this specific case in linear response, and
find that the relaxation rate shows a substantial increase as the
temperature gets closer to the critical temperature. In three
dimensions the relaxation rate does, however, remain finite at the
transition temperature.

{\it Spin drag Hall effect}--- To illustrate the spin drag Hall
effect we consider first a homogeneous three-dimensional Bose
mixture of two spin states, which we label $|0\rangle$ and
$|1\rangle$, in the normal state. We assume that the bosons in
spin state $|1\rangle$ couple to an external force ${\bf F}$, and
that the other spin state does not couple to this external force.
(Generalizations to more than two spin species and different
forces are straightforward.) This would for example be the case if
the external force is due to the Zeeman effect in a magnetic field,
and if the two spin states correspond to the $m_F=0$ and $m_F=1$
projections of an $F=1$ hyperfine state. We assume that the system
is rotating, which gives rise to a Coriolis force that is the
equivalent of the the Lorentz force from the electronic Hall
effect.

The appropriate Boltzmann equation is
\begin{equation}\label{eq:B2}
\frac{\partial f_1}{\partial t} + \left[
\frac{{\bf F}}{\hbar} + {\bf \Omega}\times{\bf k}
\right]\cdot\frac{\partial f_1}{\partial {\bf k}} =
\Gamma_{\rm coll} [f_0,f_1].
\end{equation}
Here, $f_1({\bf k}, t)$ is the distribution function for the
bosons in state $|1\rangle$. The Boltzmann equation for $f_0 ({\bf
k}, t)$ is found by replacing  $f_1 \leftrightarrow f_0$ and
setting ${\bf F}$ to zero. Furthermore, ${\bf \Omega}=\Omega {\bf
z}$ is the rotation vector, and ${\bf \Omega} \times {\bf k}$
gives the Coriolis force. The collision term, $\Gamma_{\rm
coll}[f_0,f_1]$, describes collisions of atoms with different
spin. We
will give its precise definition later on. There are of course also collisions between atoms with an
identical spin but they do not play a role for the spin drag.

We solve the Boltzmann equation by using the {\it ansatz} $f_1
({\bf k}, t) = N_B\left(\epsilon_{{\bf k} - m {\bf
v}_1(t)/\hbar}\right)$, with a similar expression for $f_0 ({\bf
k},t)$ in terms of ${\bf v}_0 (t)$. Here, $N_B (\epsilon) =
[e^{\beta (\epsilon-\mu})-1]^{-1}$ is the Bose-Einstein
distribution function with $\beta = (k_B T)^{-1}$, as usual, the
inverse thermal energy, $k_B$ Boltzmann's constant, and $T$ the
temperature. The single-particle dispersion is $\epsilon_{\bf k} =
\hbar^2 {\bf k}^2/2m$ with $m$ the particle mass. The chemical
potential $\mu$ is determined by the condition that the density of
atoms is constant. The Boltzmann equation leads to the following equations of motion
for the drift velocities ${\bf v}_0 (t)$ and ${\bf v}_1 (t)$
\begin{eqnarray}
nm\frac{d {\bf v}_ 0}{d t} &=& 2 n m
{\bf \Omega}\times{\bf v}_0 - {\bf \Gamma}({\bf v}_0 -
{\bf v}_1);\\
nm\frac{d {\bf v}_ 1}{d t} &=& n {\bf F}  + 2 n m
{\bf \Omega}\times{\bf v}_1 + {\bf \Gamma}({\bf v}_0 - {\bf v}_1).
\end{eqnarray}
Here, $n$ is the particle density per spin. We assume this density
to be equal for the two spin species. Again, note that generalizations of
the above to spin and mass imbalanced systems are straightforward.

In general the frictional spin drag is determined by the full
nonlinear (vector-valued) function ${\bf \Gamma}({\bf v}_0 - {\bf
v}_1) = \int d {\bf k}\hbar {\bf k} \Gamma_{\rm
coll}[N_B\left(\epsilon_{{\bf k} - m {\bf
v}_0/\hbar}\right),N_B\left(\epsilon_{{\bf k} - m {\bf
v}_1/\hbar}\right)]/(2\pi)^3 $. In the linear-response regime
where the velocities are small, we make use of the fact that it
can be approximated by ${\bf \Gamma}({\bf v}_0 - {\bf v}_1) \simeq
\Gamma'(0) ({\bf v}_0 - {\bf v}_1)$. Note that here we make use of
the isotropy of the collision term in the Boltzmann equation. We
now introduce ${\bf j} = n({\bf v}_1 - {\bf v}_0)$ the relative
particle current, which up to dimensionful prefactors is equal to
the spin current, and solve the above equations of motion for the
steady state, i.e., $d{\bf j}/dt = 0$. We then find that ${\bf j}
= \bm{\sigma}\cdot{\bf F} = \bm{\rho}^{-1}\cdot {\bf F}$, which defines the
conductivity and resistivity tensors $\bm{\sigma}$ and $\bm{\rho}$,
respectively. Note that these are $3\times 3$ matrices since the
force and current are three-dimensional vectors. We find that the
longitudinal resistivities $\rho_{xx} = \rho_{yy} =\rho_{zz}= 2
\Gamma'(0)/n^2$, which are related to the spin drag relaxation
time $\tau$ via a Drude-like formula as $\rho_{xx} \equiv
m/n\tau$. This relaxation time is the time scale on which the spin
current decays due to collisions of atoms in different spin states. The Hall resistivities are given by $\rho_{xy} =
-\rho_{yx} = 2 m \Omega/n$. All other components of the
resistivity and conductivity tensors are zero. Like the Hall
resistivity in electronic systems, the transverse components of
the resistivity do not depend on the specifics of the processes
that lead to a nonzero longitudinal resistivity, but only on the
density and strength of the Coriolis force. The longitudinal
resistivity, however, that determines the dissipation of the
relative momentum current via frictional spin drag, depends on the
inter-spin-species collisions.

{\it Collective modes} ---
To implement the spin drag Hall effect in a realistic cold-atom
experiment, we have to take into account the effects of the
trapping potential. Steady-state current are now no longer
possible. In this situation the collective-mode spectrum of the
mixture provides an experimental method to determine the spin drag
resistivities. We consider a harmonic trapping potential $V_{\rm
trap} ({\bf x}) = m\omega_0^2 ({x}^2+y^2)/2+m\omega_z^2z^2/2$ with
radial trapping frequency $\omega_0$ and axial frequency $\omega_z
$. We now have for the Boltzmann equation
\begin{equation}
\frac{\partial f_1}{\partial t} + [{\bf \Omega}\times{\bf k} -
\frac{1}{\hbar} \bm{\nabla} V] \cdot \frac{\partial f_1}{\partial
\bf{k}} + \frac{\hbar \bf{k}}{m}\cdot \frac{\partial
f_1}{\partial \bf{x}} = \Gamma_{\rm coll}[f_0,f_1],
\end{equation}
where $V (\bx)=V_{\rm trap} ({\bf x})-m\Omega^2(x^2+y^2)/2$ includes the centrifugal force.
The equation for $f_0$ is again found by replacing $f_1 \leftrightarrow f_0$.

We solve this inhomogeneous Boltzmann equation by making the {\it
ansatz} $ f_1({\bf x}, {\bf k}, t) = N_B (\epsilon_{{\bf k} - m
{\bf v}_1(t)/\hbar} + V({\bf x}-{\bf x}_1 (t)))$, with a similar
expression for $f_0 ({\bf x},{\bf k},t)$. This {\it ansatz} is now
parameterized by the center-of-mass velocities ${\bf v}_{0,1} (t)$
and positions ${\bf x}_{0,1} (t)$ of the two atomic clouds. From
this, we get the equations of motion.
\begin{eqnarray}\label{eq:EOM2}
\nonumber N m \frac{d{\bf v}_1}{dt} = 2N m {\bf \Omega} \times {\bf v}_1 - N
\frac{dV\!\left( {\bf x}_1\right)}{d{\bf x}_1} \\+ {\bf \Gamma}({\bf v}_0
-
{\bf v}_1, {\bf x}_0 - {\bf x}_1) ;\\
\nonumber N m \frac{d{\bf v}_0}{dt} = 2N m {\bf \Omega} \times {\bf v}_0 - N
\frac{dV\!\left( {\bf x}_0\right)}{d{\bf x}_0} \\- {\bf \Gamma}({\bf v}_0
- {\bf v}_1, {\bf x}_0 - {\bf x}_1),
\end{eqnarray}
with $N$ the particle number per spin state. Note that due to the
centrifugal force, we need to have that $|\Omega|<\omega_0$.

We again linearize the above equations using that ${\bf
\Gamma}({\bf v}, {\bf x}) \simeq \Gamma' {\bf v}$ due to the
isotropy of the collision integral.
We next observe that all the motion in the $z$-direction
decouples. We therefore only consider the motion of the clouds in
the $x-y$-plane, since this contains the spin drag Hall effect.
The linearized equations then yield a collective-mode spectrum
with eight modes in total. There are four modes in which the two
clouds of particles move in phase, and in which there is, as a
result, no drag effect. The modes correspond physically to
in-phase harmonic oscillations of the two clouds with frequencies
$\omega_0 \pm \Omega$. There are two different frequencies because
the degeneracy due to the two equivalent directions of oscillation
in the effective two-dimensional system, is split by the external
rotation. The four out-of-phase modes correspond physically to the
two atomic clouds moving relative to each other. This results in
transfer of momentum between the two clouds, leading to spin drag
and damping of these modes. These modes have the frequencies
\begin{equation}
\omega = - i \gamma \pm \Omega + \sqrt{\omega_0^2 +(i\gamma\pm\Omega)^2
}
\end{equation}
The imaginary part of the above frequencies gives the damping rate
of the modes, and is for $\Omega \ll \omega_0$ given  by $\gamma \equiv 1/2\tau=\Gamma'/Nm$ with $\tau$ the
spin drag relaxation time. This relaxation time gives the
longitudinal spin drag resistivity as
$\rho_{xx}=\rho_{yy}=m/n\tau$, and is estimated next. From the
eigenvectors of the modes we find that the plane of oscillation of
the out-of-phase dipole mode is not fixed in the co-rotating frame,
which implies a transverse spin current. This is the trap
equivalent of the spin drag Hall effect discussed in the previous
section.

 {\it Spin drag relaxation time} --- In the inhomogeneous case, we find that
 \begin{eqnarray}&&
 \hspace*{-0.5cm}\Gamma({\bf v}_0 - {\bf v}_1, {\bf x}_0 - {\bf x}_1) = \int d {\bf x} \int \frac{d \bf{k}}{(2\pi)^3} \hbar\bf{k} \nonumber \\
 && \times \Gamma_{\rm coll} \left[N_B (\epsilon_{{\bf k} - m
{\bf v}_0(t)/\hbar} + V({\bf x}\!-\!{\bf x}_0 (t))), \right. \nonumber \\
 && \hspace*{1.1cm} \left. N_B (\epsilon_{{\bf k} - m {\bf v}_1(t)/\hbar} + V({\bf
x}\!-\!{\bf x}_1 (t)))\right]~,
\end{eqnarray}
where
\begin{eqnarray*}
&& \hspace*{-0.4cm} \nonumber \Gamma_{\rm coll}[f_0,f_1] =
\frac{(2\pi)^4}{\hbar} (T^{2B}_{01})^2 \int \frac{d {\bf
k}_2}{(2\pi)^3} \int \frac{d {\bf
k}_3}{(2\pi)^3} \int \frac{d {\bf k}_4}{(2\pi)^3} \\
&&\times\delta({\bf k} + {\bf k}_2 - {\bf k}_3 - {\bf k}_4)
\delta(\epsilon_{\bf k} + \epsilon_{{\bf k}_2} - \epsilon_{{\bf
k}_3} - \epsilon_{{\bf k}_4}) \\ &&\times \{[1 + f_1({\bf x}, {\bf
k}, t)][1 + f_0({\bf x},{\bf k}_2, t)]f_1({\bf x},{\bf k}_3,
t)f_0({\bf x},{\bf k}_4, t)  \\ &&- f_1({\bf x},{\bf k},
t)f_0({\bf x},{\bf k}_2, t)[1 + f_1({\bf x}, {\bf k}_3, t)][1 +
f_0({\bf x},{\bf k}_4, t)]\}.
\end{eqnarray*}
Here, $T_{01}^{2B}$ is the two-body T-matrix, which equals $4 \pi
a \hbar^2 /m$, with $a$ the scattering length for
inter-spin-species collisions. Introducing the response function
\begin{eqnarray}
&& \chi({\bf x};{\bf q},\omega) \\ &&= \int \frac{d {\bf
k}}{(2\pi)^3} \frac{N_B(\epsilon_{\bf k}+V({\bf x})) -
N_B(\epsilon_{{\bf k}+{\bf q}}+V({\bf x}))}{\epsilon_{{\bf k}+{\bf
q}}-\epsilon_{\bf k} + \hbar\omega + i 0}, \nonumber
\end{eqnarray}
we find
\begin{eqnarray}
\Gamma' &=& \frac{\hbar^2}{12 \pi \beta} (T^{2B}_{01})^2
  \int d {\bf x} \int \frac{d {\bf q}}{(2\pi)^3} \int_{-\infty}^\infty d\omega
 q^2 \nonumber \\ &&\times~ \frac{[{\rm Im}[\chi({\bf x};{\bf q}, \omega)]]^2}{\sinh^2(\beta
 \hbar\omega/2)},
\end{eqnarray}
from which we can determine the spin drag relaxation time. The
imaginary part of the response function is worked out explicitly
to yield
\begin{eqnarray}
&&{\rm Im}[\chi({\bf x}; {\bf q}, \omega)] \\ &&=
\frac{m}{2\hbar^2\Lambda q}\log\left(\frac{e^{\frac{q^2
\Lambda^2}{16 \pi} - \beta\mu({\bf x}) -
\frac{\hbar\beta\omega}{2} + \frac{\pi(\hbar\beta\omega)^2}{q^2
\Lambda^2}} - e^{-\hbar\beta\omega}}{e^{\frac{q^2 \Lambda^2}{16
\pi} - \beta\mu({\bf x}) - \frac{\hbar\beta\omega}{2} +
\frac{\pi(\hbar\beta\omega)^2}{q^2 \Lambda^2}} - 1}\right),
\nonumber
\end{eqnarray}
with $\Lambda$ the thermal de Broglie wavelength and $\mu({\bf
x})=\mu - V({\bf x})$.

\begin{figure}
\vspace{-0.5cm} \centerline{\epsfig{figure=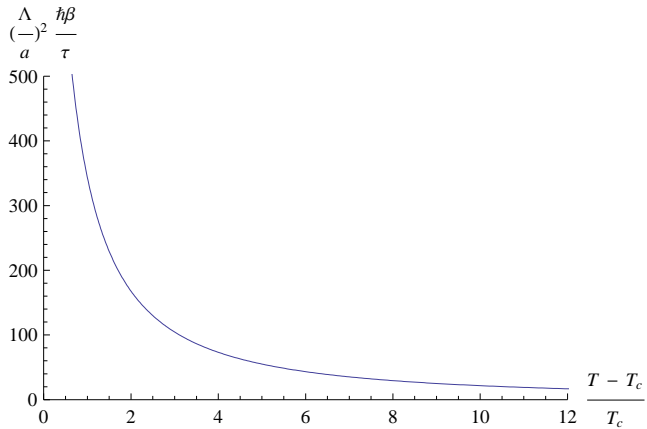}}
 \caption{Bose enhancement of the spin drag relaxation time $\tau$ upon approaching the critical temperature $T_c$ for Bose-Einstein condensation from above.}
 \label{fig:taudrag}
\end{figure}

We estimate the above expression for a three-dimensional
homogeneous system with density $n$ for which we, in first
approximation, have take the central density in the trap to make
connection with the inhomogeneous case. The result for $\tau$ is
evaluated numerically, and is shown in Fig.~\ref{fig:taudrag}.
We see that Bose enhancement is indeed at play: $1/\tau$ gets
dramatically bigger as the temperature approaches the critical
temperature. We do find, however, that it remains finite as $T
\downarrow T_c$. Furthermore, from our numerical results, we find
that $1/\tau(T) - 1/\tau(T_c) \propto - \sqrt{-\beta\mu} \propto
T_c - T$.

{\it Discussion and conclusion}--- We have introduced the spin
drag Hall effect, i.e., the emergence of a transverse spin
current, in rotating Bose mixtures. To determine whether the spin
drag relaxation rate can be measured in principle, we make
estimates of the relaxation time for some realistic values of the
parameters. Taking for instance $^{87}$Rb at temperatures between
10 and 100 nK, with an inter-species scattering length of about
100 Bohr radii, we find values of the order of $1 - 100$ ms for densities $n=10^{11}-10^{12}$ cm$^{-3}$.
Considering that the trapping potential usually has $\omega_0
\simeq 0.01 - 1$ kHz, this means the damping should indeed happen
on an observable time scale. To compare with electronic systems we note that the Drude formula $m/n e^2 \tau$ (with $e$ the electronic charge to convert to units of electrical resistivity) with our result for $\tau$ yields resistivities of the order of $10^{-6}-10^{-2}$ $\Omega$m, many orders of magnitude larger than the spin drag resistivity in an electronic system \cite{Physics.2.87}.

One interesting aspect of the Bose enhancement of the $1/\tau$ is
the behavior close to the critical temperature. Here, we
numerically found that $1/\tau(T)-1/\tau(T_c) \propto -
(T-T_c)^\kappa$, with $\kappa =1$ within our numerical accuracy.
In future work we intend to investigate the value of this exponent
in various dimensions with renormalization-group methods, taking
into account critical fluctuations that are not captured by the
Boltzmann approach presented here. Note that in our previous work
concerning one spatial dimension, we found a divergence of the
spin drag relaxation rate as $(T-T_c)^{-5/2}$
\cite{PhysRevLett.103.170401}, also with Boltzmann methods. An
interesting aspect of a two-component Bose mixture is that it also
may become ferromagnetic above the critical temperature of
Bose-Einstein condensation. We intend to study also the effects of
this transition on the spin drag.

The collective mode spectrum determined theoretically in this
Letter can be observed experimentally by setting the two spin
states in relative motion. This can for example be achieved by
shortly applying small magnetic field gradient to excite the
spin-dipole mode. Another possibility is to use a state-selective
laser in a manner that is similar to the generation of the
second-sound dipole mode in a partially Bose-Einstein condensed
gas \cite{PhysRevLett.103.265301}.

We hope that the close collaboration between theory and
experiments in this area, will lead to more insight into bosonic
transport and, on the long run, may eventually lead to the
development of useful atomtronics devices \cite{pepino2009}, where atoms rather than
electrons are the main carriers of transport.

This work
was supported by the Stichting voor Fundamenteel Onderzoek
der Materie (FOM), the Netherlands Organization
for Scientific Research (NWO), and by the European
Research Council (ERC) under the Seventh Framework
Program (FP7).



\end{document}